\documentclass[conference]{IEEEtran}
\IEEEoverridecommandlockouts
\usepackage{cite}
\usepackage{amsmath,amssymb,amsfonts}
\usepackage{algorithmic}
\usepackage{graphicx}
\usepackage{textcomp}
\usepackage[table]{xcolor}
\usepackage{makecell, cellspace}
\usepackage{multirow}

\ifCLASSOPTIONcompsoc
  \usepackage[caption=false,font=normalsize,labelfont=sf,textfont=sf]{subfig}
\else
  \usepackage[caption=false,font=footnotesize]{subfig}
\fi

\def\BibTeX{{\rm B\kern-.05em{\sc i\kern-.025em b}\kern-.08em
    T\kern-.1667em\lower.7ex\hbox{E}\kern-.125emX}}
\begin{document}

\title{From Design Principles to Prototype: A Game for Students with ADHD and Learning Disabilities Transitioning to Post-Secondary Education}

\author{
\IEEEauthorblockN{
Avery Keuben\textsuperscript{1},
Talaal Irtija\textsuperscript{1},
Joseph Tandyo\textsuperscript{1},
Stefanie Ng\textsuperscript{1},
Amy Wiebe\textsuperscript{1}
}
\IEEEauthorblockN{
Samuel Gaudet\textsuperscript{1},
Rebekah Leslie\textsuperscript{1,2},
Meadow Schroeder\textsuperscript{1},
Lauren Goegan\textsuperscript{3},
Richard Zhao\textsuperscript{1}
}
\IEEEauthorblockA{\textsuperscript{1}University of Calgary, Calgary, Alberta, Canada}
\IEEEauthorblockA{\textsuperscript{2}University of Aberdeen, Aberdeen, United Kingdom}
\IEEEauthorblockA{\textsuperscript{3}University of Manitoba, Winnipeg, Manitoba, Canada}
}



\maketitle


\begin{abstract}
Students with Attention Deficit Hyperactivity Disorder (ADHD) and Learning Disabilities (LD) can face significant academic, social, and organizational challenges when transitioning to post-secondary education. This paper presents a literature-informed serious game prototype designed to support this transition. We synthesize prior work into design considerations for students with ADHD and LD and show how these considerations are instantiated in a story-driven game.
\end{abstract}

\section{Introduction}

The transition from high school to post-secondary education requires students to develop new academic, social, organizational, and self-advocacy skills. This transition can be especially challenging for students diagnosed with Attention Deficit Hyperactivity Disorder (ADHD) and Learning Disabilities (LD) \cite{lipka_adjustment_2020}. ADHD can manifest as ``a continuous pattern of lack of attention and/or hyperactivity and/or impulsiveness'' \cite{angileri_serious_2024}, while LD is defined by ``difficulties learning and using academic skills,'' with diagnostic specifiers related to reading, mathematics, and written expression \cite{guha2014diagnostic}. Prior work has shown that students with ADHD and LD may experience lower academic and personal adjustment than their peers \cite{lipka_adjustment_2020}. Students with ADHD have also been found to be less prepared for post-secondary education, with a lack of self-determination skills, living skills, and academic readiness skills identified as possible contributing factors \cite{canu_college_2021}. For students with LD, perceived academic ability, academic integration, and social integration are important determinants of post-secondary success and satisfaction \cite{goegan_online_2022}.

Existing transition resources may not fully support students who benefit from interactive, structured, and accessible learning experiences \cite{goegan_students_2020}. Parker and Banerjee \cite{parker_leveling_2007} further found that post-secondary students with ADHD and LD had less exposure to online and technology-blended courses than their peers. These findings suggest a design opportunity for tools that help students rehearse transition-related decisions, navigate common post-secondary scenarios, and engage with institutional expectations in a low-stakes environment.

In this paper, we approach this problem as a systems design challenge: how can findings from literature on ADHD, LD, post-secondary transition, accessibility, and serious game design be translated into concrete design considerations and instantiated in a working prototype? We contribute a systems contribution of how relevant literature was synthesized into design considerations and how those considerations were operationalized in a playable game system. Specifically:
\begin{itemize}
\item We analyze prior literature to derive game design considerations tailored to the learning needs of students with ADHD and LD.
\item We present a literature-informed story-driven serious game prototype for students with ADHD and LD that simulates common scenarios transitioning to post-secondary education. 
\end{itemize}

\begin{table*}[htbp]
    \begin{center}
    \caption{Literature-derived game design considerations for students with ADHD and LD.}

    \begin{tabular}{|p{34mm}|p{58mm}|p{72mm}|}
    \hline
       \textbf{Design Consideration} & \textbf{Literature-Informed Rationale} & \textbf{Game Design Implication}\\\hline       

       \rowcolor{lightgray!50} 
       \textbf{Category: Content} & & \\\hline   

       Narrative Structure & 
       Prior game design literature suggests that fictional worlds and characters can provide structure for activities \cite{lamsa_games_2018,sonne2016assistive}, while educational game narratives can provide a reason to learn and a rationale for gameplay \cite{jemmali2018educational}. & 
       The game should use a coherent story world in which transition-related learning is embedded in character interactions, quests, and decisions, rather than presented as disconnected informational content. \\\hline

       Content Tailoring & 
       Students with ADHD may experience executive-function challenges, and routine-building can help desired behaviors become more natural over time \cite{angileri_serious_2024}. & 
       Game scenarios should model realistic transition tasks, such as responding to emails, meeting counselors, visiting campus offices, and preparing routines for post-secondary life. \\\hline

       Cognitive Load Management & 
       Prior work recommends content that is easy to learn in short sessions \cite{lamsa_games_2018}, simple activities that do not require excessive time commitment \cite{pykhtina2012designing}, and modular levels that players can explore in flexible order \cite{jemmali2018educational}. & 
       The game should divide transition preparation into short chapters, mini-games, and modular activities that allow students to progress incrementally without being overwhelmed. \\\hline

       \rowcolor{lightgray!50} 
       \textbf{Category: Engagement} & &\\\hline   

       Feedback and Guidance & 
       Prior work emphasizes positive feedback, opportunities for reflection, sufficient structure, contextual instructions, and hints to support player reasoning \cite{angileri_serious_2024,pykhtina2012designing,jemmali2018educational}. & 
       The game should provide immediate, encouraging, and context-sensitive feedback, including hints, pop-ups, and next-step guidance when students encounter uncertainty. \\\hline

       Progress and Rewards & 
       Scoring and feedback are central to game design \cite{lamsa_games_2018}; students with ADHD may benefit from immediate rewards \cite{angileri_serious_2024}; and prior work recommends praise, rewards, and clear process visibility \cite{sonne2016assistive,mcknight_designing_nodate}. & 
       The game should make progress visible through quests, badges, task lists, and milestones that reinforce completion and help players understand where they are in the transition process. \\\hline

       Adaptive Learning & 
       Prior work recommends monitoring user performance to maintain an appropriate challenge level \cite{lamsa_games_2018}, personalizing exercises \cite{angileri_serious_2024}, and using data-driven methods to identify knowledge areas that designers may overlook \cite{zhao2019knowledge}. & 
       The game should track player choices and progress so that later dialogue, tasks, or support prompts can respond to the player's demonstrated needs and prior interactions. \\\hline 

       \rowcolor{lightgray!50} 
       \textbf{Category: Interface Design} & & \\\hline   

       Minimalistic Interface & 
       ADHD-oriented design recommendations emphasize neat, uncluttered layouts \cite{mcknight_designing_nodate}, minimizing distractions \cite{sonne2016assistive}, and focusing attention on relevant stimuli \cite{angileri_serious_2024}. & 
       The interface should reduce visual clutter, provide consistent navigation, and focus attention on the current task, for example through a constrained smartphone-style interface. \\\hline

       Accessibility Features & 
       Prior work recommends allowing users to adjust appearance or functionality \cite{lamsa_games_2018}, including options such as large font sizes for users with ADHD \cite{mcknight_designing_nodate}. & 
       The game should include adjustable text, audio, visual, and interaction settings so students can personalize the experience according to their learning needs and preferences. \\\hline
    \end{tabular}
    \end{center}
    
    \label{tab:design_considerations}
\end{table*}

\section{Related Works}

Video games have been used by researchers and practitioners for a range of intervention and training purposes, including reducing loneliness for children with chronic illness \cite{alexandridis2023first} or treating chronic depression \cite{desai2010creating}. Within ADHD- and LD-related contexts, Supangan et al. \cite{supangan_gamified_2019} developed a mobile game as a supplementary tool for teaching mathematics, language, and basic hygiene to children with ADHD. Mancera et al. \cite{mancera_atenderah_2017} presented a game to train cognitive abilities related to e-learning for students with ADHD. Empowered Brain \cite{keshav_digital_2019} is an augmented reality game for students with autism spectrum disorder, with researchers showing that performance in Empowered Brain may be predictive of ADHD-related symptoms. Ahmad et al. \cite{ahmad_game-based_2010} showed that simple brainteasers and word games can produce positive feedback from students with LD. Kose et al. \cite{kose_where_2022}, however, cautioned that supervised and structured intervention may be necessary for technology-based visual perception rehabilitation to be effective. Mosher et al. \cite{mosher_step-by-step_2022} similarly emphasized that not all technology tools are suitable for students with ADHD, and developed a process for selecting appropriate technology supports. These works demonstrate the potential of games and interactive technologies for ADHD, LD, accessibility, and health-related learning contexts. However, comparatively less work has examined how serious games can be designed to support students with ADHD and LD during the transition to post-secondary education. Our work addresses this design space by synthesizing prior literature into game design considerations and implementing those considerations in a prototype focused on post-secondary transition scenarios.

\begin{figure*}[htb]
    \centering
    \subfloat[Dialogue options when conversing with a character\label{a}]{%
       \includegraphics[width=0.41\linewidth]{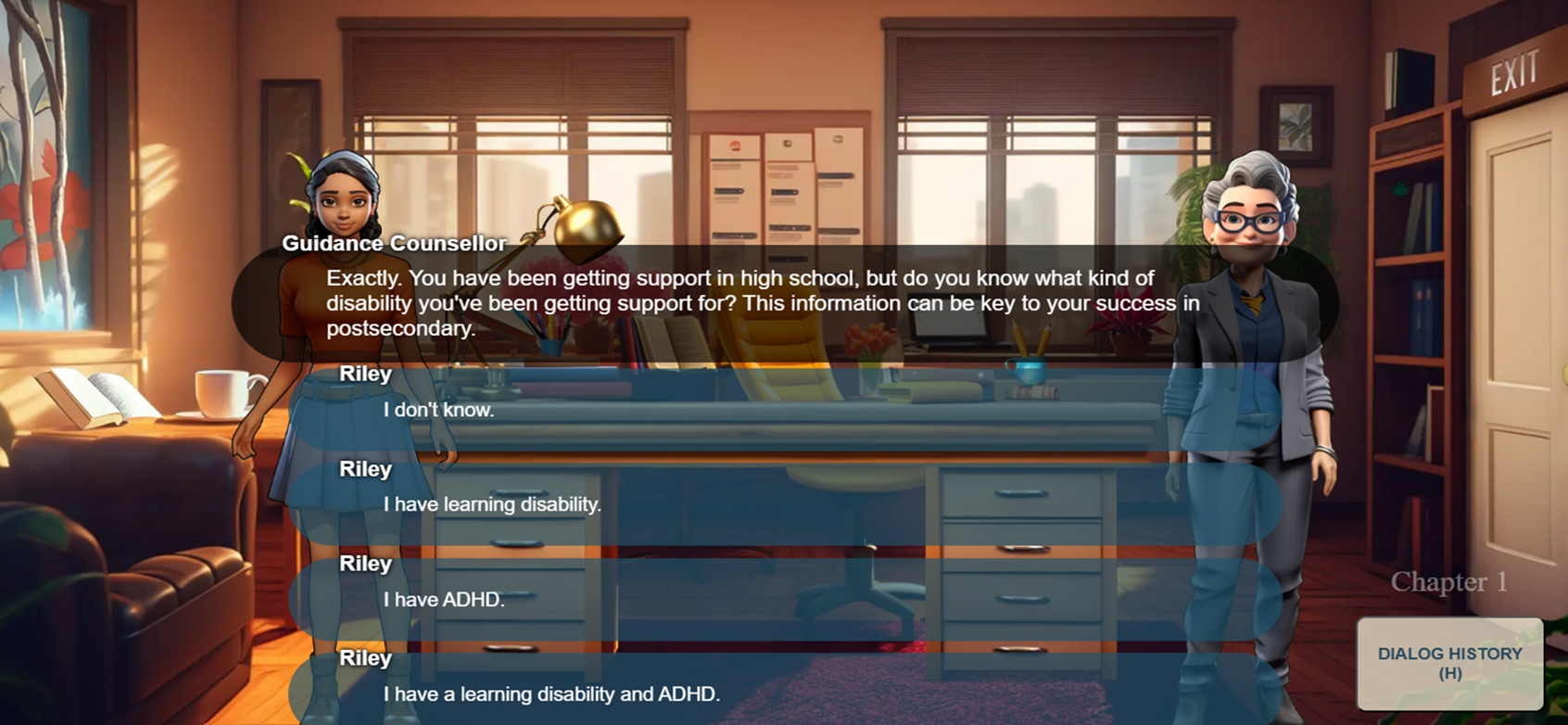}}
    \hfill
    \subfloat[``Jeopardy''-style mini-game \label{b}]{%
        \includegraphics[width=0.41\linewidth]{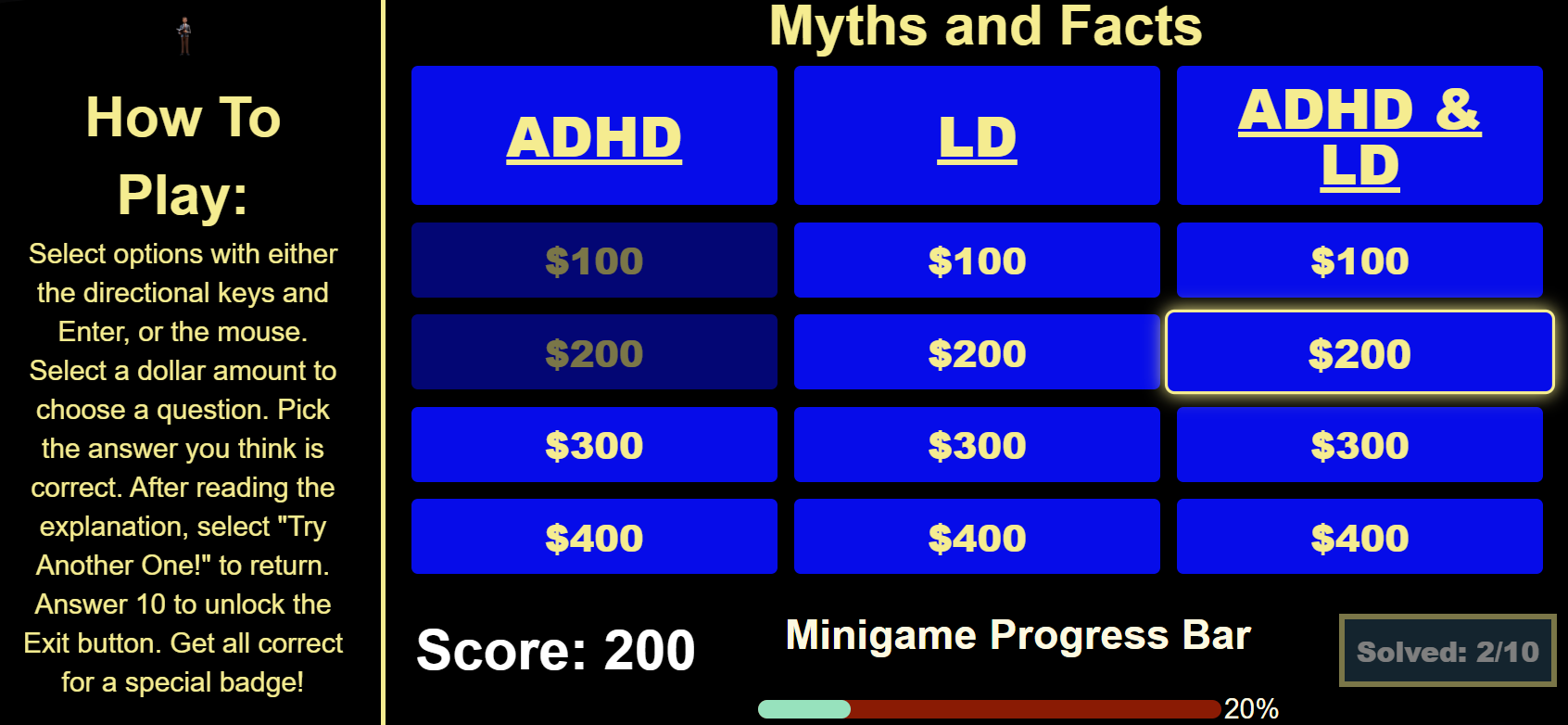}}
    
    \subfloat[In-game smartphone interface\label{c}]{%
       \includegraphics[width=0.21\linewidth]{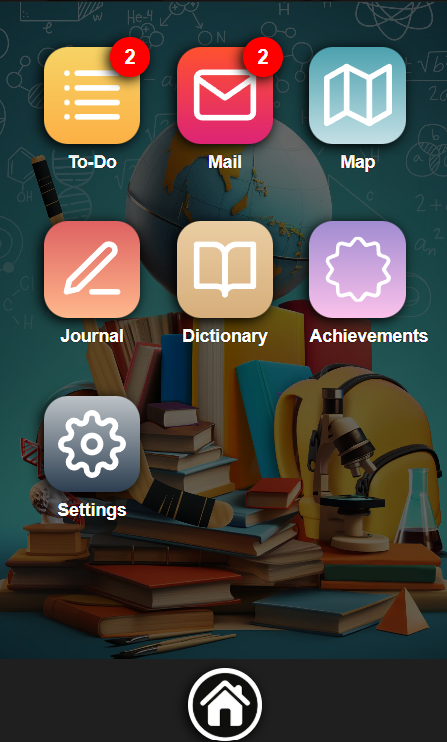}}
    \hfill
    \subfloat[Journal app showing the to-do list\label{d}]{%
        \includegraphics[width=0.21\linewidth]{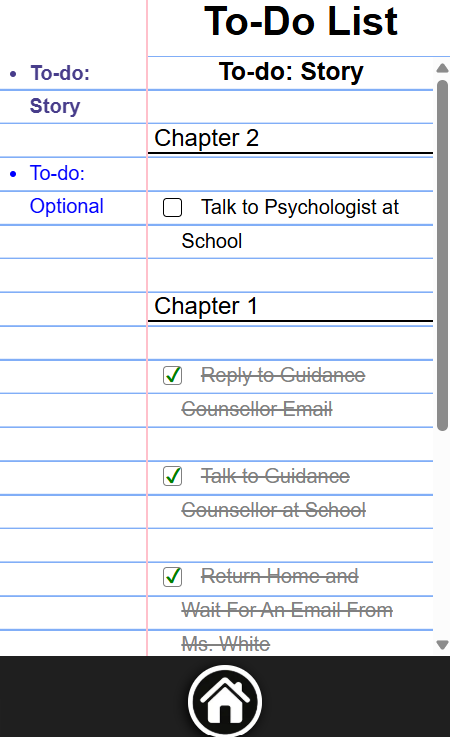}}
    \hfill
    \subfloat[Journal app showing the badges\label{e}]{%
        \includegraphics[width=0.21\linewidth]{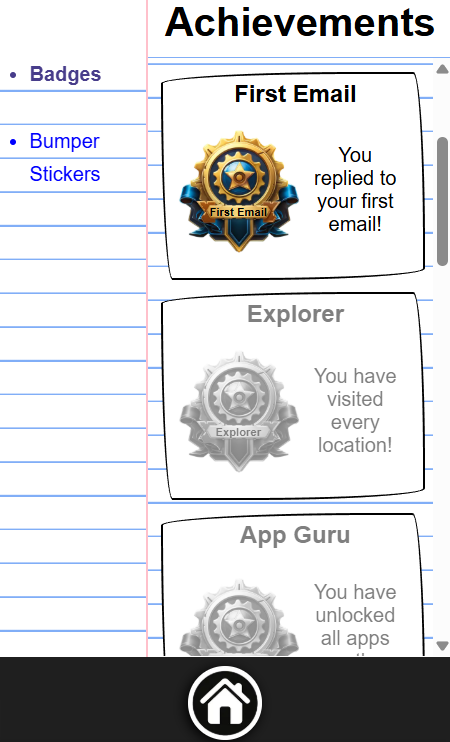}} 
    \hfill
    \subfloat[Map app showing the places to explore\label{f}]{%
        \includegraphics[width=0.21\linewidth]{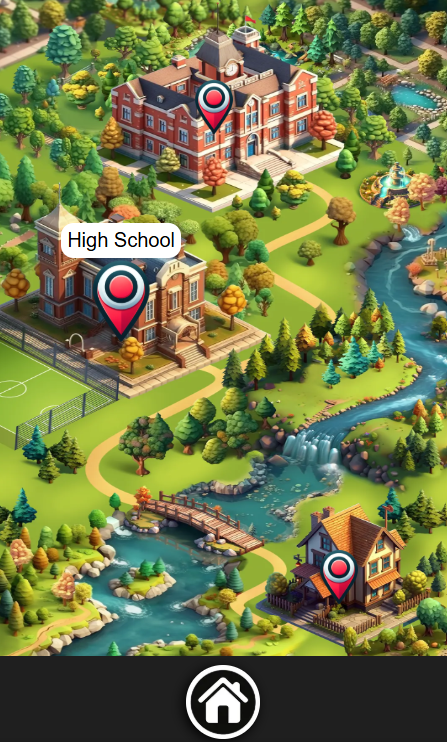}}

  \caption{Example functionalities in the game: (a) shows the ability to choose between three dialogue options; (b) shows a mini-game that serves as a fun quiz; (c) shows the in-game smartphone in which a player can access apps; (d) shows the journal app with the to-do list of items; (e) shows the badges received through game play; (f) shows the map of the game world containing the home, high school, and university.}
  \label{fig:tasks} 
\end{figure*}

\section{Literature-Derived Design Considerations for Students with ADHD and LD}

The design of serious games for students with ADHD and LD requires more than applying general accessibility or educational software recommendations. In this work, we synthesize prior literature on ADHD, LD, serious games, accessibility, and post-secondary transition support to identify design considerations specific to a game-based transition preparation system. While prior work, including McKnight \cite{mcknight_designing_nodate}, has proposed software design guidelines for children with ADHD, these recommendations are not directly framed around serious games, post-secondary transition, or the combined needs of students with ADHD and LD. Our contribution is therefore to translate and reorganize relevant findings into game-specific design considerations that can inform the design of an interactive transition-support prototype.

We analyze existing literature together with the needs of students with ADHD and LD in the context of transition to post-secondary education, aided by Universal Design for Learning \cite{UDL_2024}. Rather than treating the literature as a set of generic interface recommendations, we organize the resulting considerations around three game-design concerns: Content, Engagement, and Interface Design. The Content considerations address how transition-related knowledge can be embedded into narrative scenarios and short learning activities. The Engagement considerations address how feedback, rewards, and adaptation can support motivation and self-paced progression. The Interface Design considerations address how the system can reduce cognitive load and increase accessibility. Table~\ref{tab:design_considerations} summarizes these literature-derived design considerations and the corresponding literature that informed them.

\section{Implementing the Design Considerations in a Transition-to-Post-Secondary Game}

In this section, we detail the design process following the design considerations in developing a story-driven web-based game, GEARS\footnote{https://sites.google.com/view/gears-game}, designed to support the transition to post-secondary education for high school students with ADHD and LD. GEARS offers a self-paced, choose-your-own-adventure experience that closely mirrors real-world scenarios encountered by students during this critical period. The goal of the game is to increase preparedness by allowing students to encounter, rehearse, and reflect on transition-related tasks in a structured, low-stakes environment. We delineate below how each of the outlined considerations has been implemented within the game to ensure it serves as an effective and engaging resource for students with ADHD and LD.

\textbf{Narrative Structure}: The game casts the player as a high school student exploring post-secondary opportunities, beginning in the player's bedroom. Players can choose an avatar from a set of diverse options. The narrative progresses across educational and institutional settings, guiding the player through interactions with a guidance counselor (Figure~\ref{fig:tasks}a), psychologist, peers, and university offices, such as accessibility and housing services. The game is organized into short chapters, each focused on a specific aspect of the transition process and culminating in preparation for the student's move to post-secondary education.

\textbf{Content Tailoring}: The game scenarios are designed to reflect events and decisions that students may encounter during the transition to post-secondary education, such as receiving and responding to emails from a guidance counselor, attending university open houses, learning about accessibility services, and navigating institutional expectations. We developed the content with subject matter experts on ADHD and LD as situated gameplay rather than as isolated informational text.

\textbf{Cognitive Load Management}: Mini-games, such as a ``Jeopardy'' game show-style activity at the end of a chapter (Figure~\ref{fig:tasks}b), reinforce key concepts while providing a structured break from narrative progression. This modular structure allows players to engage with the material incrementally rather than completing long or complex tasks in a single session.

\textbf{Feedback and Guidance}: Pop-up messages appear in a context-sensitive manner on the top-right side of the screen to direct attention to relevant actions, clarify next steps, and reduce uncertainty.

\textbf{Progress and Rewards}: Badges reward both small actions, such as answering an email, and larger milestones within the transition process (Figure~\ref{fig:tasks}e). The to-do list tracks current objectives and completed tasks (Figure~\ref{fig:tasks}d), helping players understand where they are in the game and what remains to be done.

\textbf{Adaptive Learning}: Personalized progression is achieved through examples such as when a player shows hesitation in discussing ADHD, the game can introduce a peer character who encourages seeking support. This allows later dialogue and activities to respond to prior player behavior, connecting adaptive support to the player's demonstrated needs.

\textbf{Minimalistic Interface}: The interface uses clear and consistent navigation organized around an in-game smartphone. Through this smartphone, players access apps such as the journal, map, and other resources (Figure~\ref{fig:tasks}c--f). When opened, the current game area is blurred so that attention remains focused on the smartphone interface and its available actions.

\textbf{Accessibility Features}: The prototype includes personalization options through a settings app, including avatar customization, text-size adjustment, audio volume control, and background changes. Dialogue text is also converted to speech using a text-to-speech tool. These features support students with diverse learning needs and preferences while keeping accessibility integrated into the main game system.

These design considerations involve inherent trade-offs,  requiring balancing structure with player agency, simplicity with informational depth, and personalization with predictability. For example, short chapters and explicit guidance reduce cognitive demands but may constrain exploration, while adaptive dialogue can make the experience more relevant but increases authoring complexity and may make progression less transparent. Similarly, accessibility and customization options support diverse needs, but too many controls may add interface complexity. The prototype therefore prioritizes modular interaction while retaining limited choice and customization. 

\section{Conclusions and Future Work}

This paper presented the design and implementation of a story-driven serious game prototype to support students with ADHD and LD during the transition to post-secondary education. We synthesized prior literature into design considerations for this context and showed how these considerations were operationalized through narrative scenarios, modular activities, feedback, progress tracking, adaptive dialogue, a minimalistic interface, and accessibility features. We plan to examine, through longer-term user study, how game-based transition support may affect students' preparedness and engagement with post-secondary resources.

\bibliographystyle{IEEEtran}
\bibliography{paperbib}

@article{mosher_step-by-step_2022,
	title = {A {Step}-{By}-{Step} {Process} for {Selecting} {Technology} {Tools} for {Students} {With} {ADHD}},
	volume = {37},
	issn = {0162-6434, 2381-3121},
	doi = {10.1177/0162643420978570},
	language = {en},
	number = {2},
	urldate = {2024-03-02},
	journal = {Journal of Special Education Technology},
	author = {Mosher, Maggie A. and Carreon, Adam C. and Sullivan, Barbara J.},
	month = jun,
	year = {2022},
	pages = {310--317},
}

@article{keshav_digital_2019,
	title = {Digital {Attention}-{Related} {Augmented}-{Reality} {Game}: {Significant} {Correlation} between {Student} {Game} {Performance} and {Validated} {Clinical} {Measures} of {Attention}-{Deficit}/{Hyperactivity} {Disorder} ({ADHD})},
	volume = {6},
	issn = {2227-9067},
	shorttitle = {Digital {Attention}-{Related} {Augmented}-{Reality} {Game}},
	doi = {10.3390/children6060072},
	language = {en},
	number = {6},
	urldate = {2024-03-02},
	journal = {Children},
	author = {Keshav, Neha U. and Vogt-Lowell, Kevin and Vahabzadeh, Arshya and Sahin, Ned T.},
	month = may,
	year = {2019},
	pages = {72},
}

@inproceedings{supangan_gamified_2019,
	address = {Singapore Singapore},
	title = {A gamified learning app for children with {ADHD}},
	isbn = {978-1-4503-6092-0},
	doi = {10.1145/3313950.3313966},
	language = {en},
	urldate = {2024-03-02},
	booktitle = {Proceedings of the 2nd {International} {Conference} on {Image} and {Graphics} {Processing}},
	publisher = {ACM},
	author = {Supangan, Renz Anthony and Acosta, Leo Alfred S. and Amarado, Jose Lorenzo S. and Blancaflor, Eric B. and Samonte, Mary Jane C.},
	month = feb,
	year = {2019},
	pages = {47--51},
}

@inproceedings{mancera_atenderah_2017,
	address = {Timisoara, Romania},
	title = {{aTenDerAH}: {A} {Videogame} to {Support} e-{Learning} {Students} with {ADHD}},
	isbn = {978-1-5386-3870-5},
	shorttitle = {{aTenDerAH}},
	doi = {10.1109/ICALT.2017.157},
	language = {en},
	urldate = {2024-03-02},
	booktitle = {2017 {IEEE} 17th {International} {Conference} on {Advanced} {Learning} {Technologies} ({ICALT})},
	publisher = {IEEE},
	author = {Mancera, Laura and Baldiris, Silvia and Fabregat, Ramon and Gomez, Sergio and Mejia, Carolina},
	month = jul,
	year = {2017},
	pages = {438--440},
}

@article{parker_leveling_2007,
	title = {Leveling the {Digital} {Playing} {Field}: {Assessing} the {Learning} {Technology} {Needs} of {College}-{Bound} {Students} {With} {LD} and/or {ADHD}},
	volume = {33},
	issn = {1534-5084, 1938-7458},
	shorttitle = {Leveling the {Digital} {Playing} {Field}},
	doi = {10.1177/15345084070330010201},
	language = {en},
	number = {1},
	urldate = {2024-03-03},
	journal = {Assessment for Effective Intervention},
	author = {Parker, David R. and Banerjee, Manju},
	month = oct,
	year = {2007},
	pages = {5--14},
}

@article{kose_where_2022,
	title = {Where {Exactly} {Is} the {Therapist} in {Virtual} {Reality} and {Game}-{Based} {Rehabilitation} {Applications}? {A} {Randomized} {Controlled} {Trial} in {Children} with {Specific} {Learning} {Disability}},
	volume = {11},
	issn = {2161-783X, 2161-7856},
	shorttitle = {Where {Exactly} {Is} the {Therapist} in {Virtual} {Reality} and {Game}-{Based} {Rehabilitation} {Applications}?},
	doi = {10.1089/g4h.2021.0241},
	language = {en},
	number = {3},
	urldate = {2024-03-03},
	journal = {Games for Health Journal},
	author = {Köse, Barkın and Temizkan, Ege and Aran, Orkun Tahir and Galipoğlu, Hasan and Torpil, Berkan and Pekçetin, Serkan and Karabulut, Erdem and Şahin, Sedef},
	month = jun,
	year = {2022},
	pages = {200--206},
}

@article{lamsa_games_2018,
	title = {Games for enhancing basic reading and maths skills: {A} systematic review of educational game design in supporting learning by people with learning disabilities},
	volume = {49},
	issn = {0007-1013, 1467-8535},
	shorttitle = {Games for enhancing basic reading and maths skills},
	doi = {10.1111/bjet.12639},
	language = {en},
	number = {4},
	urldate = {2024-03-03},
	journal = {British Journal of Educational Technology},
	author = {Lämsä, Joni and Hämäläinen, Raija and Aro, Mikko and Koskimaa, Raine and Äyrämö, Sanna‐Mari},
	month = jul,
	year = {2018},
	pages = {596--607},
}

@inproceedings{ahmad_game-based_2010,
	address = {Kuala Lumpur, Malaysia},
	title = {Game-based learning courseware for children with learning disabilities},
	isbn = {978-1-4244-6715-0},
	doi = {10.1109/ITSIM.2010.5561303},
	language = {en},
	urldate = {2024-03-03},
	booktitle = {2010 {International} {Symposium} on {Information} {Technology}},
	publisher = {IEEE},
	author = {Ahmad, Wan Fatimah Wan and Akhir, Emelia Akashah P. and Azmee, Sarah},
	month = jun,
	year = {2010},
	pages = {1--4},
}

@article{goegan_online_2022,
	title = {Online {Learning} for {Students} with {Learning} {Disabilities} and {Their} {Typical} {Peers}: {The} {Association} between {Basic} {Psychological} {Needs} and {Outcomes}},
	volume = {37},
	issn = {0938-8982, 1540-5826},
	shorttitle = {Online {Learning} for {Students} with {Learning} {Disabilities} and {Their} {Typical} {Peers}},
	doi = {10.1111/ldrp.12277},
	language = {en},
	number = {2},
	urldate = {2024-03-03},
	journal = {Learning Disabilities Research \& Practice},
	author = {Goegan, Lauren D. and Daniels, Lia M.},
	month = may,
	year = {2022},
	pages = {140--150},
}

@article{goegan_students_2020,
	title = {Students with {LD} at {Postsecondary}: {Supporting} {Success} and the {Role} of {Student} {Characteristics} and {Integration}},
	volume = {35},
	issn = {0938-8982, 1540-5826},
	shorttitle = {Students with {LD} at {Postsecondary}},
	doi = {10.1111/ldrp.12212},
	language = {en},
	number = {1},
	urldate = {2024-03-03},
	journal = {Learning Disabilities Research \& Practice},
	author = {Goegan, Lauren D. and Daniels, Lia M.},
	month = feb,
	year = {2020},
	pages = {45--56},
}

@article{canu_college_2021,
	title = {College {Readiness}: {Differences} {Between} {First}-{Year} {Undergraduates} {With} and {Without} {ADHD}},
	volume = {54},
	issn = {0022-2194, 1538-4780},
	shorttitle = {College {Readiness}},
	doi = {10.1177/0022219420972693},
	language = {en},
	number = {6},
	urldate = {2024-03-03},
	journal = {Journal of Learning Disabilities},
	author = {Canu, Will H. and Stevens, Anne E. and Ranson, Loren and Lefler, Elizabeth K. and LaCount, Patrick and Serrano, Judah W. and Willcutt, Erik and Hartung, Cynthia M.},
	month = nov,
	year = {2021},
	pages = {403--411},
}

@article{lipka_adjustment_2020,
	title = {Adjustment to {Higher} {Education}: {A} {Comparison} of {Students} {With} and {Without} {Disabilities}},
	volume = {11},
	issn = {1664-1078},
	shorttitle = {Adjustment to {Higher} {Education}},
	doi = {10.3389/fpsyg.2020.00923},
	language = {en},
	urldate = {2024-03-03},
	journal = {Frontiers in Psychology},
	author = {Lipka, Orly and Sarid, Miriam and Aharoni Zorach, Inbar and Bufman, Adi and Hagag, Adi Anna and Peretz, Hila},
	month = jun,
	year = {2020},
	pages = {923},
}

@article{mcknight_designing_nodate,
	title = {Designing for {ADHD}: in search of guidelines},
	language = {en},
  author={McKnight, Lorna},
  journal={IDC 2010 Digital Technologies and Marginalized Youth Workshop},
  year={2010}
}

@article{angileri_serious_2024,
	title = {A serious web game for children with attentive disorders: design and experiences from two trials},
	volume = {39},
	issn = {0737-0024, 1532-7051},
	shorttitle = {A serious web game for children with attentive disorders},
	doi = {10.1080/07370024.2023.2240797},
	language = {en},
	number = {1-2},
	urldate = {2024-03-04},
	journal = {Human–Computer Interaction},
	author = {Angileri, Letizia and Manca, Marco and Paternò, Fabio and Santoro, Carmen},
	month = mar,
	year = {2024},
	pages = {43--78},
}

@inproceedings{alexandridis2023first,
  title={First Experiments with an Applied Gaming Intervention for reducing Loneliness of Children with Chronic Illness: Lessons Learned},
  author={Alexandridis, Dionysis and Bakkes, Sander CJ and Nijhof, Sanne L and Van De Putte, Elise and Veltkamp, Remco C},
  booktitle={Proceedings of the 18th International Conference on the Foundations of Digital Games},
  pages={1--11},
  year={2023}

}

@inproceedings{desai2010creating,
  title={Creating video games to treat chronic depression},
  author={Desai, Neesha and Szafron, Duane and Sayegh, Liliane and Greiner, Russell and Turecki, Gustavo},
  booktitle={GRAND Annual Conference},
  year={2010}
}

@inproceedings{zhao2019knowledge,
  title={Knowledge assessment: game for assessment of symptoms of child physical abuse},
  author={Zhao, Richard and Shelton, Christopher R and Hetzel-Riggin, Melanie D and LaRiccia, Jordan and Louchart, Gregory and Meanor, Adam and Risser, Heather J},
  booktitle={Proceedings of the 14th International Conference on the Foundations of Digital Games},
  pages={1--7},
  year={2019}
}

@inproceedings{jemmali2018educational,
  title={Educational game design: an empirical study of the effects of narrative},
  author={Jemmali, Chaima and Bunian, Sara and Mambretti, Andrea and El-Nasr, Magy Seif},
  booktitle={Proceedings of the 13th International Conference on the Foundations of Digital Games},
  pages={1--10},
  year={2018}
}

@misc{guha2014diagnostic,
  title={Diagnostic and statistical manual of mental disorders: DSM-5},
  author={Guha, Martin},
  year={2014},
  publisher={Emerald Group Publishing Limited}
}

@misc{UDL_2024, 
  url={https://udlguidelines.cast.org/}, 
  journal={UDL}, 
  year={2018},
  title={Universal Design for Learning Guidelines version 2.2},
  author={CAST}
}

@inproceedings{sonne2016assistive,
  title={An assistive technology design framework for ADHD},
  author={Sonne, Tobias and Marshall, Paul and Obel, Carsten and Thomsen, Per Hove and Gr{\o}nb{\ae}k, Kaj},
  booktitle={Proceedings of the 28th Australian conference on computer-human interaction},
  pages={60--70},
  year={2016}
}

@article{pykhtina2012designing,
  title={Designing for attention deficit hyperactivity disorder in play therapy: the case of Magic Land},
  author={Pykhtina, Olga and Balaam, Madeline and Wood, Gavin and Pattison, Sue and Olivier, Patrick},
  journal={Journal of Internet Psychology},
  volume={4},
  year={2012}
}
\end{document}